\documentstyle[amsmath,amssymb,amsfonts,12pt]{elsart}

\def\text#1{\mbox{#1}}

\def\ee{\end{equation}}
\def\be#1{\begin{equation}\label{#1}}

\def\ba{\begin{array}}
\def\ea{\end{array}}

\begin{document}

\begin{frontmatter}
\title{{Gauge hierarchy from a topological viewpoint?}}
\author{M. O. Tahim, C. A. S. Almeida}
\address{Universidade Federal do Cear\'{a}\\ Physics Department\\ C.P. 6030, 60470-455 Fortaleza-Ce,
Brazil\thanksref{e-mail}}
\thanks[e-mail]{Electronic addresses: mktahim@fisica.ufc.br, carlos@fisica.ufc.br}

\begin{abstract}
In this work we explore an alternative to the central point of the
Randall-Sundrum brane world scenario, namely, the particular
nonfactorizable metric, in order to solve the hierarchy problem.
From a topological viewpoint, we show that the exponential factor,
crucial in the Randall-Sundrum model, appears in our approach,
only due to the brane existence instead of a special metric
background. Our results are based in a topological gravity theory
via a non-standard interaction between scalar and non-abelian
degrees of freedom and in calculations about localized modes of
matter fields on the brane. We point out that we obtain the same
results of the Randall-Sundrum model using only one 3-brane, since
a specific choice of a background metric is no longer required.
\end{abstract}

\end{frontmatter}

PACS: 11.10.Kk, 04.50.+h, 12.60.-i, 04.20.Gz

Keywords: Hierarchy problem; Randall-Sundrum model; Topological
theory.

\vspace{1.0cm} May the Standard Model be placed in form of the
recent insights coming from String Theories, where several
dimensions appear so naturally? The standard model for strong,
weak and electromagnetic interactions, described by the gauge
group $SU(3)\times SU(2)\times U(1)$, has its success strongly
based on experimental evidences. However, it has several serious
theoretical drawbacks suggesting the existence of new and
unexpected physical facts beyond those discussed in the last
years. One of these problems is the so called \textit{gauge
hierarchy problem} which is related to the weak and Planck scales,
the fundamental scales of the model. The central idea of this
problem is to explain the smallness and radiative stability of the
hierarchy $M_{ew}/ M_{pl}\sim 10^{-17}$. In the context of the
minimal standard model,  this hierarchy of scales is unnatural
since it requires a fine-tuning order by order in the perturbation
theory. The first attempts to solve this problem were  the
technicolor scenario \cite{A} and the low energy supersymmetry
\cite{B}.

With the string theories, the search of many-dimensional theories
became important. The basic idea is that extra dimensions can be
used to solve the hierarchy problem: the fields of the standard
model must be confined to a $(3+1)$-dimensional subspace, embedded
in a $n$-dimensional manifold. In the seminal works of
Arkani-Hamed, Dimopoulos, Dvali and Antoniadis \cite{C}, the
$4$-dimensional Planck mass is related to $M$, the fundamental
scale of the theory, by the extra-dimensions geometry. Through the
Gauss law, they have found $M^{2}_{pl}=M^{n+2} V_{n}$, where
$V_{n}$ is the extra dimensions volume. If $V_{n}$ is large
enough, $M$ can be of the order of the weak scale. However, unless
there are many extra dimensions, a new hierarchy is introduced
between the compactification scale, $\mu_{c}= V^{-\frac{1}{n}}$,
and $M$. An important feature of this model is that the space-time
metric is factorizable, i.e., the $n$-dimensional space-time
manifold is approximately a product of a $3$-dimensional space by
a compact $(n-3)$-dimensional manifold.

Because of this new hierarchy, Randall and Sundrum \cite{D} have
proposed a higher dimensional scenario that does not require large
extra dimensions, neither the supposition of a metric factorizable
manifold. Working with a single $S^{1}/Z_{2}$ orbifold extra
dimension, with three-branes of opposite tensions localized on the
fixed points of the orbifold and with adequate cosmological
constants as $5$-dimensional sources of gravity, they have shown
that the space-time metric of this model contains a redshift
factor which depends exponentially on the radius $r_{c}$ of the
compactified dimension:
\begin{equation}\label{eq1}
d s^{2}= e^{-2k r_{c}|\phi|}\eta_{\mu\nu} d x^{\mu}d x^{\nu}-r_{c}
d\phi^{2},
\end{equation}
where $k$ is a parameter of the order of $M$, $x^{\mu}$ are
Lorentz coordinates on the surfaces of constant $\phi$, and
$-\pi\leq\phi\leq\pi$ with $(x,\phi)$ and $(x,-\phi)$ identified.
The two $3$-branes are localized on $\phi=\pi$ and $\phi=0$. In
fact, this scenario is well known in the context of string theory
\cite{E}. The non-factorizable geometry showed in Eq.(\ref{eq1})
has at least two important consequences that will be discussed
here. The first one is that the $4$-dimensional Planck mass is
given in terms of the fundamental scale $M$ by
\begin{equation}\label{eq2}
M_{pl}^{2}=\frac{M^{3}}{k}[1-e^{-2k r_{c}\pi}],
\end{equation}
in such a way that, even for large $k r_{c}$, $M_{pl}$ is of the
order of $M$. The second one is that because of the exponential
factor on the space-time metric, a field confined to a $3$-brane
at $\phi=\pi$ with mass parameter $m_{0}$ will have physical mass
$m_{0}e^{-k r_{c}\pi}$ and for $k r_{c}$ near of $12$, the weak
scale is dynamically generated by the fundamental scale $M$ which
is of the order of the Planck mass.

On the other hand, background independent theories are welcome. As
an example it is worth mentioning the Quantum Loop Gravity,
developed mainly by Asthekar et al. \cite{asthekar,asthekar1}.
Also the problem of background dependence of string field theory
has not been successfully addressed. The string field theory has a
theoretical problem: it is only consistently quantized in
particular backgrounds, which means that we have to specify a
metric background in order to write down the field equations of
the theory. This problem is fundamental because a unified
description of all string backgrounds would make possible to
answer questions about the selection of particular string vacua
and in general to give us a more complete understanding of
geometrical aspects of string theory \cite{witten}.

In this work we explore an alternative to the central point of the
Randall-Sundrum model, namely, the particular nonfactorizable
metric. Using a topological theory, we show that the exponential
factor, crucial in the Randall-Sundrum model, appears in our
approach, only due to the brane existence instead of a special
metric background.

Some searches have been made trying to implement branes as
topological defects in order to solve the hierarchy problem
\cite{Wolfe}. Here the brane is simulated by a $3$-dimensional
domain wall embedded in a $5$-dimensional space-time. Domain walls
are simple solitons, objects whose great stability is due to the
nontrivial topology of the parameter space of the theory \cite{F}.
They only appear after phase transitions, specifically, when
discrete symmetries are broken.

In order to study the hierarchy problem we choose to work with
topological gravity. Motivated by current searches in the quantum
gravity context \cite{smolin,asthekar}, we study topological
gravity of $B\wedge F$ type \cite{Diakonov,K}. Then, we can affirm
that our model is purely topological because $1)$ the brane exists
due to the topology of the parameter space of the model and $2)$
gravity is metric independent. We will see that these features
give us interesting results when compared to the Randall-Sundrum
model.


The model is based on the following action:
\begin{equation}\label{eq3}
S= \int d^{5} x
[\frac{1}{2}\partial_{\mu}\theta\partial^{\mu}\theta+
k\varepsilon_{\mu\nu\alpha\rho\lambda}\theta H_{\mu\nu\alpha}^{a}
F_{\rho\lambda}^{a}-V(\theta)].
\end{equation}
In this action the $\theta$ field is a real scalar field that is
related to the domain wall. In this context, the presence of a
kinetic term for the $\theta$ field (together with the symmetry
breaking potential), is required to construct a topological defect
(the domain wall). We remark that the $\theta$ field acts as a
background field in order to provide a brane where we have an
effective BF-type theory. The fields $H_{\mu\nu\alpha}^{a}$ and
$F_{\rho\lambda}^{a}$ are non-abelian gauge fields strengths and
will be related to the gravitational degrees of freedom. Namely, in
pure gauge theory,
$H^{a}_{\mu\nu\alpha}=\partial_{\mu}B^{a}_{\nu\alpha}-\partial_{\nu}B^{a}_{\alpha\mu}-\partial_{\alpha}B^{a}_{\mu\nu}+gf^{abc}A^{b}_{\mu}B^{c}_{\nu\alpha}$
and
$F^{a}_{\mu\nu}=\partial_{\mu}A^{a}_{\nu}-\partial_{\nu}A^{a}_{\mu}+g'f^{abc}A^{b}_{\mu}A^{c}_{\nu}$.
The second term of this action is a topological term that
generalizes to $D=5$ the theta-term of QCD. To see this, it is
enough to do a simple dimensional reduction, namely, define $B_{\mu
5}^{a}=-B_{5 \mu}^{a}=V^{\mu}_{a}$, $A_{5}^{a}=\varphi$,
$\varepsilon_{5\nu\alpha\rho\lambda}\equiv\varepsilon_{\nu\alpha\rho\lambda}$
and $\partial_{5} G(x^{\mu})=0$, where $G$ is any field of this
model. Then, the theta-term arises as a result from the
compactification procedure defined above, as

\begin{equation}
\int d^{5} x k\varepsilon_{\mu\nu\alpha\rho\lambda}\theta
H_{\mu\nu\alpha}^{a} F_{\rho\lambda}^{a}\rightarrow \int d^{4} x
k'\varepsilon_{\nu\alpha\rho\lambda}\theta V_{\nu\alpha}^{a}
F_{\rho\lambda}^{a},
\end{equation}

where $V_{\nu\alpha}^{a}=\partial_{\nu}V_{\alpha}^{a}-
\partial_{\alpha}V_{\nu}^{a}+g f^{abc}V_{\nu}^{b} V_{\alpha}^{c}$.
Identifying $V_{\alpha}^{a}$ with $A_{\alpha}^{a}$ we obtain the
term discussed. Because of this fact, the $\theta$ field can be
thought as the axionic field. The axion has appeared as a proposal
to solve the strong CP problem \cite{H}. The presence of
instantons in the theory results in an effective term added to the
QCD action, namely, $\sim\int d^{4} x
\varepsilon^{\nu\alpha\rho\lambda}\theta F_{\nu\alpha}^{a}
F_{\rho\lambda}^{a}$, which violates CP symmetry. The problem is
solved when  we add to the theory the axionic field with the
imposition of a new symmetry, the Peccei-Quinn symmetry, that is
$\theta\rightarrow\theta+\textit{a}$ (\textit{a} is a constant
which contains the CP violating quantities of the theory). The
action (\ref{eq3}) is invariant under the Peccei-Quinn symmetry
transformation

\begin{equation}
\theta\rightarrow\theta+2\pi n.
\end{equation}

The axionic potential is

\begin{equation}\label{eq4}
V(\theta)=\lambda(1-\cos\theta),
\end{equation}

which preserves the Peccei-Quinn symmetry. Nevertheless, it is
spontaneously broken in scales of the order of $M_{PQ}\sim
10^{10}-10^{12}$GeV. This value is obtained from cosmological and
experimental constraints \cite{I}. The potential (\ref{eq4}) is
not interesting for our purposes. The fact is that domain walls
appeared for the first time in the Universe in the QCD phase
transition era, i.e., when $T_{QCD}\sim 100 MeV$ \cite{J}, a scale
relatively close to the weak scale $M_{ew}\sim 10^{3} GeV$. In
this situation, the Peccei-Quinn symmetry is explicitly broken
$(U_{PQ}(1)\rightarrow Z(N))$ by instanton effects. It is possible
to simulate this explicit break by a simple theoretical field toy
model. For such, we write $V(\theta)$ as a polinomial potential in
powers of $\theta$, what is equivalent to take terms only up to
the second order in the expansion of the Eq. (\ref{eq4}). We
propose the following potential
\begin{equation}\label{eq5}
V(\theta)=\frac{\lambda}{4}(\theta^{2}-v^{2})^{2},
\end{equation}
which explicitly breaks the $U_{PQ}(1)$ Peccei-Quinn symmetry, in
order to generate a brane in an energy close to the weak scale.
With this particular choice of the potential, the existence of the
brane is put on more consistent grounds. In other words, the brane
appears almost exactly in an energy scale of the universe near the
symmetry breaking scale of the electroweak theory. This feature
was assumed in previous works without a careful justification.
However, this mechanism leads to a large disparity between the
Planck mass $M_{PL}\sim 10^{18} GeV$ and the scale of explicit
breaking of $U_{PQ}(1)$ which is relatively close to the weak
scale, $M_{ew}\sim 10^{3} GeV$: we assume this disparity as a new
version of the hierarchy problem.

The equation of motion of the $\theta$ field considering the
potential (\ref{eq5}) is the following:

\begin{equation}\label{eq6}
\theta+\lambda\theta^{3}-\lambda
v^{2}\theta=k\varepsilon_{\mu\nu\alpha\rho\lambda}
H_{\mu\nu\alpha}^{a} F_{\rho\lambda}^{a}.
\end{equation}

This equation is easily solved. Supposing a static configuration
and that $\theta\equiv \theta(x_{4})$, the solution is:

\begin{equation}\label{eq7}
\theta(x_{4})=v\tanh(\sqrt{\frac{\lambda}{2}}v x_{4}).
\end{equation}

This solution defines a $3$-brane embedded in a
$(4+1)$-dimensional space-time. The mass scale of this model is
$m=\sqrt{\lambda}v$ and the domain wall-brane thickness is
$m^{-1}$. With this information we can now discuss the effective
theory on the domain wall-brane. An integration by parts of the
topological term in the action (\ref{eq3}) will result in
\begin{equation}\label{eq8}
\varepsilon_{\mu\nu\alpha\rho\lambda}\theta(x_{4})
H_{\mu\nu\alpha}^{a}
F_{\rho\lambda}^{a}=-3\varepsilon_{\mu\nu\alpha\rho\lambda}\partial_{\mu}\theta
B_{\nu\alpha}^{a} F_{\rho\lambda}^{a}+\ldots,
\end{equation}
where we do not consider complicated interactions and linear terms
on $\theta$ (the function (\ref{eq7}) is odd). Because of
$\theta\equiv \theta(x_{4})$ the summation on the $\mu$ index will
result only in a derivative term of the $x_{4}$ coordinate. Then,
the Levi-Civita tensor $\varepsilon_{\mu\nu\alpha\rho\lambda}$
will be an authentic four-dimensional tensor:
$\varepsilon_{4\nu\alpha\rho\lambda}\equiv
\varepsilon_{\nu\alpha\rho\lambda}$. We have assumed that the
tensors $B_{\mu\nu}^{a}$ and $A_{\rho a}$ are weakly dependent on
the $x_{4}$ coordinate. Then, the second term of the action
(\ref{eq3}) is rewritten as

\begin{equation}\label{eq9}
S\sim \int d^{4} x \varepsilon_{\nu\alpha\rho\lambda}
B_{\nu\alpha}^{a} F_{\rho\lambda}^{a} [\lim_{r_{c}\rightarrow
+\infty} k'\int_{0}^{r_{c}} d x_{4} \partial_{4}\theta(x_{4})],
\end{equation}

where $r_{c}$ represents the extra dimension. This last conclusion
denotes the domain wall-brane contribution to the effective
four-dimensional theory. We can see that, effectively on the domain
wall-brane, the theory is purely $4$-dimensional (this is important)
and is described by a non-abelian topological $B\wedge F$ term. The
importance of this fact is that there are several approaches to
topological gravity by means of $B\wedge F$ type models in $D=4$ and
by Chern-Simons models in $D=3$. In the ref. \cite{Diakonov}, the
authors construct a $SU(2)$, $D=4$ BF gravity in a basis independent
formulation. The point we would like to comment on that article is
that the tensorial field B is a 1-form gauge valued field. We stress
that the structure of the BF term in our work is the same as in the
ref. \cite{Diakonov}, i.e., our BF gravity on the brane is of the
type $SU(2)$, $D=4$.

Note that this approach opens the possibility to implement
topological gravity on the brane. In these models, the fundamental
fields are known. For example, the tetrad fields in $D=4$: the
metric is, by itself, a secondary object. The gauge symmetries of
these theories are, actually, the symmetries of the general
relativity \cite{K}. It can be shown that, under parameterizations
by tetrad fields, a $B\wedge F$ type action gives us
\begin{equation}\label{eq9a}
\int d^{4} x k\varepsilon^{\nu\alpha\rho\lambda} B_{\nu\alpha}^{a}
F_{\rho\lambda}^{a}\rightarrow k\int d^{4} x \sqrt{g}R,
\end{equation}
which is the Einstein-Hilbert action for the gravitational field,
where $R$ is the scalar curvature and $g$ stands for the space-time
metric \cite{Diakonov}. It is not well understood if Eq. (\ref{eq9})
can really describe the dynamics of the gravitational field
\cite{L}. In a model like this, the constant $k$ has a direct
relation with the Planck mass. From Eqs. (\ref{eq9}) and
(\ref{eq9a}), we can see the relation between the Planck mass
$k_{4}$ in $D=4$ and the extra dimension:
\begin{equation}\label{eq10}
k_{4}=\lim_{r_{c}\rightarrow +\infty} k'\int_{0}^{r_{c}} d x_{4}
\partial_{4}\theta(x_{4}).
\end{equation}
The limit $r_{c}\rightarrow +\infty$ ensures the topological
stability of the domain wall-brane. By the substitution of Eq.
(\ref{eq7}) in Eq. (\ref{eq10}), considering a finite $r_{c}$
(which means that the domain wall-brane is a finite object), we
can show that
\begin{equation}\label{eq11}
k_{4}=k'v(1-e^{-2y}) (1+e^{-2y})^{-1},
\end{equation}
where $y=\sqrt{\frac{\lambda}{2}}v r_{c}$ is the scaled extra
dimension. This result is very interesting: as our model is a
topological one, the exponential factor must not appear from any
special metric. Here, the exponential factor appears only due to
the domain wall-brane existence. As in the Randall-Sundrum model,
even for the large limit $r_{c}\rightarrow +\infty$, the
$4$-dimensional Planck mass has a specific value. This is the
reason why we believe that our model can be used to treat the
hierarchy problem.

We can make an estimative of the order of the extra dimension
considering that the domain wall-brane thickness is of the order
of $M_{ew}\sim 10^{3} GeV$. This means that the fields confined to
the domain wall-brane do not perceive the extra dimension, unless
they interact with energies greater than $M_{ew}$. In this case,
they can escape out of the brane, living in the higher dimensional
space-time \cite{M}. By the calculation of domain wall-brane
energy per unity volume $\sigma$ we can find a simple polynomial
equation of third degree in the $z=\theta(r_{c})$ variable,
containing all phase transition information:
\begin{equation}\label{eq12}
m^{-1}\sigma=\sqrt{2}v z-\sqrt{2} m^{-1} z^{3}.
\end{equation}
For the case of the Randall-Sundrum model, the extra dimension is
calculated through the normalized radial oscillation field
(refered by some authors as radion field \cite{arkani2}), i.e., it
is stabilized by a mechanism of symmetry breaking involving bulk
fields \cite{goldberger}.

We will now discuss about matter confined to the brane. It is a
well known fact that domain walls may have bound states of fields
attached to them \cite{M}. For the case of scalar fields, it was
shown using WKB approximation that a particular zero-mode living
in the domain wall-brane is given by the following field:
\begin{equation}\label{eq13}
\varphi'(x^{0}, \pmb{x},x^{4})=\frac{d\varphi(x^{4})}{d x^{4}}
exp({-i\pmb{k}\cdot\pmb{x}+iE x^{0}}) ; \qquad
E^{2}=(\pmb{k}\cdot\pmb{k})^{2}.
\end{equation}
In the last equation, $\frac{d\varphi(x^{4})}{d x^{4}}=C
e^{-2Ax_{4}}(1+e^{-Ax_{4}})^{-2}$, $C$ and $A$ are constant
parameters. In particular, a similar result is true for fermions.
Then the zero-modes, bosonic or fermionic ones, are scaled by an
exponential factor, just like in the Randall-Sundrum scenario.
Despite the fact that they are non-massive fields, there are
mechanisms involving several interacting fields \cite{N} that
generate spontaneous symmetry breaking in the defect core. In this
way, the confined fields can acquire non-zero masses. In order to
show this for the case of scalar fields, we use two real scalar
fields: $\phi(x^{0},\pmb{x},x^{4})$ and $\eta(x^{0},\pmb{x})$. We
regard the first one as a 4-dimensional confined field, i. e.,
$\phi(x^{0},\pmb{x},x^{4})=f(x^{4})\varphi(x^{0},\pmb{x})$, where
$f(x^{4})$ is just the warp factor that comes from the extra
dimension. The second one is a massless and purely 4-dimensional
field. We built the following lagrangian density
\begin{equation}\label{eq16}
\textsl{L}=\frac{1}{2}\partial_{\mu}\eta\partial^{\mu}\eta+\frac{1}{2}\partial_{\mu}\phi\partial^{\mu}\phi
-g\phi^{2}\eta^{2}-V(\phi),
\end{equation}
where $V(\phi)=-m^{2}\phi^{2}+\frac{\lambda}{4!}\phi^{4}$ is a
potential that spontaneously breaks the $\phi\rightarrow -\phi$
symmetry. In this case, if the extra dimension is finite and
constant then, during the phase transition, only the $\varphi$
field will oscillate, i. e., $\phi=f(x^{4})\varphi\rightarrow
f(x^{4})[v+\chi]$, where $v$ is the vacuum expectation value of
the $\phi$ field and $\chi$ is the fluctuation around the vacua.
Working out this idea in the last lagrangian we can show that,
after the phase transition, the $\eta$ field will acquire a mass
of the order of $f(x^{4})v\sim e^{-2Ax_{4}}(1+e^{-Ax_{4}})^{-1}
v$. This expression is analogous to the Randall-Sundrum result
\cite{D}, which provides a physical mass for fields of the
Standard Model corrected by the warp factor. Therefore, this
simple mechanism allows us to generate scales from fields confined
to a domain wall-brane, without the requirement of a particular
metric.

There is a final remark about gravity in this context: the matter
zero-modes live effectively in $D=4$ and, then, they must
contribute to the effective four-dimensional energy-momentum
tensor. They are, in fact, gravitational sources in the domain
wall-brane space-time. Consequently, we can construct a
propagation term for the gravitational field in $D=4$ (on the
brane). However, as can be seen from Eq. (\ref{eq9a}) it is
possible to build a propagation term for gravity from a
topological term. Therefore it is interesting to discuss if we can
use Eq. (\ref{eq9}) as an authentic propagation term for these
gravitational degrees of freedom. This will be discussed in a
forthcoming paper \cite{O}.


Summarizing, we have shown that a simple topological model in
field theory has the necessary features to solve the Gauge
Hierarchy problem in a very similar way to the one found by L.
Randall and R. Sundrum. With this model we have built a stable
$3$-brane (a domain wall-brane) that simulates our
four-dimensional Universe and we have argued the possibility of
topological gravity localization. Because of these facts, the
exponential factor appears only due to the existence of the domain
wall-brane and not from a special metric. Then, we have calculated
the effective Planck mass in $D=4$, pointing out the great
similarity between our result and that of the Randall-Sundrum
model. We have calculated a polinomial equation for the size of
the extra dimension using some features of models containing
domain walls. Finally, we have made a commentary about the
zero-modes bounded by the domain wall-brane, remarking the fact
that they are scaled by an exponential factor. This information
makes possible the emergence of the electroweak scale.

We did not comment about how to introduce the cosmological
constant in this model. In fact, in the Randall-Sundrum model the
cosmological constant is extremely important because it is
responsible for the final form of the metric given by Eq.
(\ref{eq1}). Another interesting fact is that brane models can
answer the following question: \textit{why is the cosmological
constant so small?}. These are good problems for future
investigations in this topological approach.

The analysis of models containing several domain walls is also
interesting. In this case, the potential that implements the phase
transition has various stable vacua. Domain walls will appear
interpolating these vacua in well defined positions: the distance
between two domain walls is constant due to the topological
stability of the model. Can we see this as another possible way to
solve the \textit{moduli stabilization problem}?

By virtue of the simplicity of this model, we can extend it to
include supersymmetry. Indeed, brane world models suggests
alternative mechanisms to the breaking of supersymmetry in our
universe. All of these subjects are interesting research
objectives.

The authors would like to thank Funda\c{c}\~{a}o Cearense de apoio
ao Desenvolvimento Cient\'{\i}fico e Tecnol\'{o}gico (FUNCAP) and
Conselho Nacional de Desenvolvimento Cient\'{\i}fico e
Tecnol\'{o}gico (CNPq) for financial support.


\begin{thebibliography}{99}
\addcontentsline{toc}{chapter}{Bibliografia}

\bibitem{A} L. Susskind, {\em Phys. Rev.} {\bf D20} , 2619 (1979);
S. Weinberg, {\em Phys. Rev.} {\bf D13}, 974 (1976); {\bf D19},
1277 (1979).

\bibitem{B} E. Witten, {\em Nucl. Phys.} {\bf B188}, 513 (1981).

\bibitem{C} N. Arkani-Hamedi, S. Dimopoulos and G. Dvali, {\em Phys. Lett.}
{\bf B429}, 263 (1998); I. Antoniadis, N. Arkani-Hamedi, S.
Dimopoulos and G. Dvali, {\em Phys. Lett.} {\bf B436}, 257 (1998).

\bibitem{D} L. Randall and R. Sundrum, {\em Phys. Rev. Lett.} {\bf 83}, 3370 (1999); L. Randall
and R. Sundrum, {\em Phys. Rev. Lett.} {\bf 83}, 4690 (1999).

\bibitem{E} P. Horava and E. Witten, {\em Nucl. Phys.} {\bf
B460}, 506 (1996); {\bf B475}, 94 (1996); E. Witten, {\em Nucl.
Phys.} {\bf B471}, 135 (1996).

\bibitem{asthekar}A. Ashtekar, {\em Phys. Rev. Lett.} {\bf 57}, 2224 (1986); C. Rovelli, {\em Phys. Rev.} {\bf D52}, 5743
(1995).

\bibitem{asthekar1} A. Ashtekar, J. D. Romano, R. S. Tate, {\em Phys. Rev.} {\bf
D40}, 2572 (1989).

\bibitem{witten} E. Witten, {\em Phys. Rev.} {\bf D46}, 5467 (1992).

\bibitem{Wolfe} O. DeWolfe, D. Freedman, S. S. Gubser, and A. Karch, {\em Phys. Rev.} {\bf
D62}, 046008 (2000).

\bibitem{F} R. Rajaraman, {\it Solitons and Instantons,} (North-Holland,
Amsterdam, 1982); A. Vilenkin and E. P. S. Shellard, {\it Cosmic
Strings and Other Topological Defects,}, Cambridge University
Press (1994).

\bibitem{smolin}L. Smolin, {\em J. Math. Phys.} {\bf 36}, 6417 (1995).

\bibitem{Diakonov} D. Diakonov and V. Petrov, {\em Grav.Cosmol.} {\bf 8}, 33
(2002).

\bibitem{K} D. Birmingham, M. Blau, M. Rakowski and G. Thompson,
{\it Topological Field Theory,}, {\em Phys. Rep.} {\bf 209},
(1991).

\bibitem{G} C. G. Callan, R. F. Dashen and D. J. Gross, {\em Phys.
Lett.} {\bf B63}, 334 (1976); C. Rebbi and R. Jackiw, {\em Phys.
Rev. Lett.} {\bf 37}, 172 (1976).

\bibitem{H} R. D. Peccei and H. R. Quinn, {\em Phys. Rev. Lett.}
{\bf 38}, 1440 (1977); S. Weinberg, {\em Phys. Rev. Lett.} {\bf
40}, 223 (1978).

\bibitem{I} M. C. Huang and P. Sikivie, {\em Phys. Rev.} {\bf
D32}, 1560 (1985).

\bibitem{J} M. N. Forbes and A. Zhitnitsky, {\em Phys. Rev.
Lett.} {\bf 85}, 5268 (2000); P. Sikivie, {\em Phys. Rev. Lett.}
{\bf 48}, 1156 (1982).

\bibitem{L} K. Lee, {\em Phys. Rev.} {\bf D35}, 2286 (1987).

\bibitem{M} V. A. Rubakov and N. E. Shaposhnikov, {\em Phys. Lett.} {\bf
B125}, 136 (1982).

\bibitem{arkani2} N. Arkani-Hamedi, S. Dimopoulos and G. Dvali, {\em Phys. Rev.}
{\bf D63}, 064020 (2001).

\bibitem{goldberger} W. D. Goldberger and M. B. Wise, {\em Phys.
Rev. Lett.} {\bf 83}, 4922 (1999).

\bibitem{N} D. Bazeia and R. F. Ribeiro, {\em Phys. Rev.} {\bf
D54}, 1852 (1996).

\bibitem{O} C. A. S. Almeida and M. O. Tahim, {\it in preparation}.

\end{thebibliography}
\end{document}